\documentclass[prd,twocolumn,showpacs,epsfig,preprintnumbers,amsmath,amssymb]{revtex4}

\usepackage{color}
\usepackage{epsfig,subfigure}
\DeclareGraphicsExtensions{.jpg,.pdf,.eps}

\def \bea{\begin{eqnarray}}
\def \eea{\end{eqnarray}}

\begin{document}

\title{Interior design of a two-dimensional semiclassical black hole: Quantum transition across the singularity}
\author{Dana Levanony  and Amos Ori}
\affiliation{Department of Physics, Technion, Haifa 32000, Israel}
\date{\today}
\begin{abstract}

We study the internal structure of a two-dimensional dilatonic
evaporating black hole, based on the CGHS model. At the
semiclassical level, a (weak) spacelike singularity was previously found to
develop inside the black hole. We employ here a simplified quantum
formulation of spacetime dynamics in the neighborhood of this
singularity, using a minisuperspace-like approach. Quantum
evolution is found to be regular and
well-defined at the semiclassical singularity. A
well-localized initial wave-packet propagating towards the singularity bounces off the latter and
retains its well-localized form. Our simplified quantum treatment thus
suggests that spacetime may extend semiclassically beyond the
singularity, and also signifies the specific extension.

\end{abstract}
\maketitle

\section{Introduction}\label{sec:Intro}

The inevitable occurrence of spacetime singularities inside black holes (BHs) constitutes
one of the greatest challenges in understanding the final fate of gravitational collapse.
A spacetime singularity is typically a region in spacetime where the curvature blows up.
The presence of spacetime singularities inside BHs has been established by several theorems \cite{Hawking_book},
 and is also demonstrated by numerous examples of exact BH solutions.

Such curvature singularities are often regarded in the literature (at least formally)
as the boundary of spacetime. It is not obvious that this assertion is fully justified
from the physical view-point (especially in the case of weak singularities; see below).
This formal point of view is nevertheless pragmatic, because even if spacetime does physically
extend beyond the singularity, in most cases it is unclear what this extension is: The singularity
makes the mathematical extension ambiguous, as the field equations become ill-defined there. One may hope,
however, that once the full theory of Quantum Gravity is formulated,
it will somehow resolve the spacetime singularities and thereby clarify how physics extends beyond them.

The remarkable revelation of Hawking radiation \cite{Hawking} and black-hole evaporation brought
new and intense interest into black-hole structure, and particularly into the associated information puzzle.
The latter appears to be intimately related to the presence of a singularity inside the BH: As long as the
hypersurface of singularity is treated as a boundary of spacetime, all bits of information which
encounter this hypersurface are quite inevitably "destroyed" there. For this reason one may
naturally hope that the information puzzle will eventually be resolved by Quantum Gravity eliminating spacetime singularities.

Two decades ago Callan, Giddings, Harvey and Strominger (CGHS) \cite{CGHS} introduced a toy model
for investigating the formation and evaporation of black holes. Their toy model consists of
two-dimensional gravity coupled to a dilaton field. In this two-dimensional framework the
renormalized stress-energy tensor $\hat T_{\alpha \beta}$ may be expressed in a simple explicit
form, for any prescribed background metric. This property makes the CGHS model a powerful tool
for analyzing BH evaporation. At the classical level (ignoring the semiclassical corrections),
one finds a one-parameter family of vacuum solutions, analogous to the Schwarzschild solution
in four dimensions. Once the semiclassical effects are added, the BH emits a thermal Hawking
radiation and evaporates.

Originally it was hoped that the CGHS model of BH formation and
evaporation will be free of any singularity, which would make it
an ideal model for addressing the information puzzle. However,
Russo, Susskind and Thorlacius \cite{Singularity_first_notice} soon
demonstrated that a spacelike curvature singularity actually forms
inside the semiclassical CGHS BH. This singularity was recently
analyzed in some detail \cite{Dana_Amos_interior_design}, based on
the homogeneous approximation. Quite remarkably, the singularity
turns out to be {\it weak} in Tipler's terminology
\cite{tipler_singularity_strength}
\cite{Amos_Singularity_Strength}. This means that extended
physical objects arrive at the singular hypersurface intact (as
opposed to a strong singularity where the object is totally torn
apart). What diverges at the CGHS singularity is the {\it rate of
change} of the metric and the dilaton (e.g. in terms of the proper
time along a timelike geodesic). In other words, the dilaton and
metric are continuous but non-smooth at the singularity.

Owing to the weakness of the semiclassical CGHS singularity, one may be tempted to assume
that semiclassical spacetime will somehow extend beyond the singular hypersurface.
This leads one to consider the issue of existence and uniqueness of solutions to the
semiclassical CGHS field equations, in the spacetime region laying at the future side
of the spacelike singularity---solutions which continuously  match the data coming from
the past side. It is not difficult to see that existence is not a problem, but
uniqueness fails here because we had to give up on smoothness. In fact, there is an
infinite set of such continuous semiclassical extensions beyond the singularity.
Semiclassical theory alone thus fails to predict the correct extension---even if
such an extension does exist in reality. One may hope, however, that a fully-quantized
version of the CGHS model will address this problem (e.g. by resolving the singularity).
It should tell us if spacetime indeed extends semiclassically beyond the singularity,
and if so---what the correct extension is.

Motivated primarily by the BH information puzzle, Ashtekar,
Taveras and Varadarajan (ATV) \cite{Ashtekar} recently proposed a
quantized version of the CGHS model. In their model (roughly
speaking) the dilaton and the metric conformal factor are elevated
to quantum fields defined on an $\Re^{2}$ manifold. These quantum
fields are represented by operators, within the Heisenberg
picture. The field equations thus turn into a nonlinear system of
operator partial differential equations, with initial data
specified at past null infinity. A key question underlying this
model is whether the quantum evolution, formulated in this way,
retains its regularity at the (would-be) semiclassical
singularity. It is conceivable that regularity is preserved at the
quantum level, though this still needs to be shown. If regularity
is indeed preserved, then in principle quantum evolution could be
followed within this model all the way from the collapse to full
evaporation. In particular this would clarify the state of the
quantum fields accessible to an observer at future null infinity,
thereby resolving the information puzzle. Unfortunately the
nonlinear operator equations underlying the ATV model are
extremely hard to solve. To overcome this difficulty, ATV
introduced two useful approximates: "Bootstrapping", which may be
applied to the early stage of evaporation; and the mean-field
approximation, which presumably holds near future null infinity.

In the ATV scenario, the singularity which was present in the
semiclassical picture is now replaced by a region of strong
quantum fluctuations. Quantum dynamics is presumably regular
there, which allows for a well-defined evolution of (quantum)
spacetime across this region. A-priory there are several
possibilities: (i) After a narrow strongly-fluctuating transition
region (marking the locus of the semiclassical spacelike
singularity), the fluctuations weaken and spacetime retains its
semiclassical character; or, alternatively, (ii) quantum
fluctuations never decline, and spacetime remains
strongly-fluctuating in the entire causal future of the
semiclassical singularity. Intermediate options are also possible:
For example, spacetime may become semiclassical again only in the
neighborhood of future null infinity.

Our main goal in this paper is to explore the quantum transition
across the semiclassical spacelike singularity, and to reveal the
nature of the spacetime region at the immediate neighborhood of
this singularity. Particularly we would like to find out if
spacetime indeed becomes semiclassical again after crossing the
singularity [option (i) above], and if so---what is the
semiclassical solution which takes place there. Unfortunately,
none of the approximate solutions constructed by ATV apply to this
portion of the manifold (and the operator field equations are
extremely difficult to solve). In order to circumvent this
difficulty, in this paper we shall apply one further
simplification to the quantum ATV dynamics, and treat it at the
"minisuperspace" level---that is, we shall ignore all degrees of
freedom associated with inhomogeneities.

Our motivation comes from the observation that the semiclassical
dynamics inside a macroscopic evaporating CGHS BH---and in
particular the approach to the spacelike singularity---may well be
approximated (locally) by homogeneous solutions of the CGHS field
equations (see discussion in \cite{Dana_Amos_interior_design}). In
this sense the BH interior resembles an approximately-homogeneous
cosmology, which is amendable to a quantum treatment at the
minisuperspace level. Our quantization strategy thus proceeds as
follows: (i) We employ the homogenous approximation for the BH
interior already at the semiclassical level, erasing any reference
to spatial dependence. Correspondingly we regard our variables
(metric and dilaton) as mere time-dependent variables, rather than
fields; (ii) We explore the leading-order asymptotic behavior of
our variables on approaching the spacelike singularity. (iii) We
derive an effective Hamiltonian which reproduces this
leading-order dynamics near the singularity. [Note that stages
(i-iii), which are implemented at the semiclassical level, were
already carried out in Ref. \cite{Dana_Amos_interior_design}.]
(iv) We now quantize our variables, but we do this in a way one
quantizes time-dependent variables (rather than fields). Thus, the
dynamics is now represented by a time-dependent wave-function
$\Psi$, and the effective Hamiltonian becomes a differential
operator which determines the time evolution of $\Psi$ via a
Schrodinger-type equation.

A key objective of this paper is to explore the properties of
this simplified quantum system. We first obtain the stationary
eigenfunctions, construct from them a well-localized initial wave
packet, and follow its time evolution after it hits the
singularity.

\textit{Summary of main results} - A key result in our analysis is
that {\it the quantum evolution is regular and hence well-defined}. This is not a
trivial result, since the Hamiltonian operator $\hat H $ itself is in fact
singular at the locus of the semiclassical singularity. Yet, all
its eigenfunctions are perfectly regular there, yielding a unique
regular quantum evolution across the semiclassical singularity.

Our initial conditions correspond to a well-localized wave packet which propagates
towards the singularity. At this initial stage the wave packet follows a semiclassical
orbit, as expected. As the wave packet approaches the singularity, it slightly spreads
and develops numerous wiggles. Subsequently, the wave packet bounces off the singularity,
and retains its original well-localized character. In this final stage, too,
the wave packet is found to follow a semiclassical orbit---a particular member
of the infinite set of possible (continuous) semiclassical solutions beyond the
singularity. Thus, this quantum model determines the specific semiclassical
extension which takes place to the future of the semiclassical singularity. The
extension of the dilaton turns out to
be (at the leading order near the singularity) a
time-reflection of the evolution prior to the singularity. However,
the extension of the other degree of freedom (the metric conformal factor) does
not admit the same time-reflection symmetry.

Concluding, our quantum model suggests that in an evaporating CGHS BH,
spacetime physics extends beyond the spacelike singularity by means of a well-defined
semiclassical spacetime. It remains to explore the detailed structure of this new
patch of spacetime. In this paper we only obtained its leading-order asymptotic behavior
just beyond the singularity. The extension beyond leading order is discussed in sec. VII.

We must recall, however, that the quantum model considered here is a very simplified one.
To obtain more firm and trustable results, one must resort to a fuller quantum-gravitational theory of spacetime dynamics.

The paper is organized as follows: In section II we briefly review
the main results of the semiclassical analysis carried out in Ref.
\cite{Dana_Amos_interior_design}, which are relevant to the
present work. In particular this includes the leading-order
semiclassical asymptotic behavior near the singularity, and the
corresponding effective Hamiltonian.

Section III describes our quantization procedure, and also presents the eigenfunctions
of the stationary Schrodinger equation. Section IV will be devoted to the construction
of a well-localized initial wave-packet, while Sec. V will discuss the time evolution
of the wave packet and its properties as it hits the singularity and afterward.
In Sec. VI we present the (leading-order) semiclassical extension of
spacetime beyond the singularity, as determined from the orbit of the well-localized wave
packet after it bounces off the singularity. Finally, Sec. VII is dedicated to discussion
of our results and their significance.

Throughout this paper we use general-relativistic units $c=G=1$.

\section{Background}

In Ref. \cite{Dana_Amos_interior_design} we investigated the interior of a
macroscopic evaporating CGHS BH, and particularly the singularity
which develops inside the BH. Here we shall briefly outline the main
results, and establish the  notation relevant to the quantum analysis below.

\subsection{The model and field equations}

The semiclassical CGHS model \cite{CGHS} describes gravity in
$1+1$ dimensions coupled to a dilaton $\phi $. It also includes a
cosmological constant $\lambda ^{2}$ and a large number $N>>1$ of
identical scalar matter fields $f_{i}$. We express the metric in
double-null coordinates $u,v$, namely $ds^{2}=-e^{2\rho }dudv$.
The action is \bea &\frac{{1}}{\pi }\int {dudv}[ {e^{-2\phi
}\left( {-2\rho ,_{uv}+4\ \phi ,_{u}\phi ,_{v}-\lambda
^{2}e^{2\rho }}\right) -\frac{1}{2}}\\
\nonumber
&\times\sum\limits_{i=1}^{N}{f_{i},_{u}f_{i},_{v}}+\frac{N}{{12}}\rho
,_{u}\rho ,_{v}] . \label{action}\eea The last term represents the
semiclassical correction. For notational simplicity we set
$\lambda =1$ [this choice amounts to a trivial shift $\rho
\rightarrow \rho +\ln (\lambda )$].

For convenience we introduce new variables: $\tilde{R}\equiv e^{-2\phi }$ (which was denoted $R$
in \cite{Dana_Amos_interior_design}) and
$S\equiv 2(\rho -\phi )$. The evolution equations then read
\begin{equation}
\tilde{R},_{uv}=-e^{S}-K\rho ,_{uv}\ ,\ \ \ S,_{uv}=K\rho ,_{uv}/\tilde{R},
\label{evolution_eq_mixed}
\end{equation}%
where $K\equiv N/12$, and $\rho =\frac{1}{2}(S-\ln \tilde{R})$ is to be
substituted. The constraint equations are
\bea
\tilde{R},_{ww}-\tilde{R},_{w}S,_{w}+\hat{T}_{ww}=0,
\label{constraint}\eea
where $w$ stands for either $u$ or $v$, and $\hat{T}_{ww}$ denotes the semiclassical energy fluxes (whose explicit form is not needed here).
We have set here $f_{i}=0$
(a trivial solution of the scalar field equation $f_{i,uv}=0$),
as we are dealing here with the evaporation of the BH rather than its formation.

It is useful to re-express the system of evolution equations (\ref%
{evolution_eq_mixed}) in its standard form, in which $\tilde{R},_{uv}$ and $%
S,_{uv}$ are explicitly given in terms of lower-order derivatives:%
\begin{eqnarray}
&&\tilde{R},_{uv}=-e^{S}\frac{{\left( {2\tilde{R}-K}\right) }}{{2\left( {%
\tilde{R}-K}\right) }}-\tilde{R},_{u}\tilde{R},_{v}\frac{K}{{2\tilde{R}%
\left( {\tilde{R}-K}\right) }},  \label{final_eq_simplified} \\
&&S,_{uv}=e^{S}\frac{K}{{2\tilde{R}\left( {\tilde{R}-K}\right) }}+\tilde{R}%
,_{u}\tilde{R},_{v}\frac{K}{{2\tilde{R}^{2}\left( {\tilde{R}-K}\right) }}.
\nonumber
\end{eqnarray}%
This form makes it obvious that the evolution equations become singular at $%
\tilde{R}=K$ and $\tilde{R}=0$.

\subsection{homogenous set-up}

\label{sec. Homogenous_equation} In Ref. \cite{Dana_Amos_interior_design} it
was established that the spacetime inside a macroscopic evaporating CGHS
black hole is (locally) approximately homogeneous. Motivated by this
observation, we shall restrict our attention to homogeneous solutions of the
field equations. Namely, we consider solutions which only depend on  $t\equiv v+u$.
In the homogeneous framework the
evolution equations (\ref{final_eq_simplified}) become
\begin{equation}
\ddot{\tilde{R}}=-e^{S}\frac{{\left( {2\tilde{R}-K}\right) }}{{2\left( {%
\tilde{R}-K}\right) }}-\dot{\tilde{R}}^{2}\frac{K}{{2\tilde{R}\left( {\tilde{%
R}-K}\right) }},
\label{Rdotdot}
\end{equation}
\begin{equation}
\ddot{S}=e^{S}\frac{K}{{2\tilde{R}\left( {\tilde{R}-K}\right) }}+\dot{\tilde{%
R}}^{2}\frac{K}{{2\tilde{R}^{2}\left( {\tilde{R}-K}\right) }},
\label{Sdotdot}
\end{equation}
where an over-dot denotes differentiation with respect to $t$. The
constraint equations (\ref{constraint}) now turn into a single equation
\[
\ddot{\tilde{R}}-\dot{\tilde{R}}\dot{S}+\hat{T}=0,
\]%
where $\hat{T}\equiv \hat{T}_{vv}=\hat{T}_{uu}$.

\subsection{The $\tilde R=K$ Singularity}

The semiclassical evolution equations are singular at $\tilde{R}=0$ and $%
\tilde{R}=K$. In a black-hole interior $\tilde{R}$ decreases monotonically,
starting from a macroscopic value ($\tilde{R}>>K$) at the horizon.
Therefore, of these two singular values, it is  $\tilde{R}=K$ which is first
encountered. Since the evolution beyond this singularity is apriory unknown,
it is not clear whether the semiclassical $\tilde{R}=0$ singularity
will ever be encountered. For this
reason we concentrate here on the $\tilde{R}=K$ singularity (which is
\textit{guaranteed} to develop).

As it turns out, this singularity is characterized by the divergence of $%
\dot{\tilde{R}}$ and $\dot{S}$, while $\tilde{R}$ and $S$ are finite. The
evolution equations are dominated near the singularity by \cite%
{Dana_Amos_interior_design}
\begin{equation}
\ddot{\tilde{R}}=-\frac{\dot{\tilde{R}}^{2}}{2(\tilde{R}-K)}\ ,\ \ \ \ddot{S}%
=\frac{\dot{\tilde{R}}^{2}}{2K(\tilde{R}-K)}.  \label{IS_equations}
\end{equation}%
The general solution is
\begin{equation}
\tilde{R}\left( t\right) =K+B\left\vert {t-t_{0}}\right\vert ^{\frac{2}{3}},
\label{Solution_at_K_R}
\end{equation}%
\begin{equation}
S\left( t\right) =-\frac{B}{K}\left\vert {t-t_{0}}\right\vert ^{\frac{2}{3}%
}+(t-t_{0})B_{2}+B_{3}.  \label{Solution_at_K_S}
\end{equation}%
It depends on four arbitrary parameters $B,B_{2},B_{3},t_{0}$ as required.
We find that $\tilde{R}$ and $S$ are indeed continuous at ${t=t_{0}}$, but $%
\dot{\tilde{R}}$ and $\dot{S}$ diverge as $|t-t_{0}|^{-1/3}$.

To simplify the description of the asymptotic behavior near $\tilde{R}=K$ we
re-define our variables as
\[
R\equiv \tilde{R}-K\ ,\ \ \ Z\equiv \tilde{R}+KS.
\]%
The above near-singularity evolution equations then decouple and take the simple form
\begin{equation}
\ddot{R}=-\frac{\dot{R}^{2}}{2R}\ ,\ \ \ \ddot{Z}=0.  \label{RZ_dotdot}
\end{equation}

\subsection{Effective Hamiltonian for the near-singularity region}\label{SubSec:effect_H}

Next we construct an effective Hamiltonian which yields the leading-order
asymptotic behavior near $\tilde{R}=K$, with the aim of subsequently
quantizing this dynamical system.

For the $R$-motion one finds the Hamiltonian \cite{Dana_Amos_interior_design}
\begin{equation}
H=\frac{P^{2}}{R}  \label{Hamiltonian_Classic}
\end{equation}%
(up to an arbitrary multiplicative constant), where $P$\ is the momentum
conjugate to $R$. For $Z$\ we obviously have the free-particle Hamiltonian $%
H_{Z}=\frac{1}{2}P_{Z}^{2}$, where $P_{Z}$\ is the momentum conjugate to $Z$%
. The overall leading-order effective Hamiltonian is $H+H_{Z}$.

Note that the effective
Hamiltonian (\ref{Hamiltonian_Classic}) is independent of $t$, hence $H$ must be conserved
along any semiclassical solution $R(t)$. Indeed one finds that $\dot
{R}=\partial H/\partial p=2P/R$ and therefore Eq. (\ref{Solution_at_K_R})
yields a constant $H$,
\bea
H=\frac{R\dot{R}^{2}}{4}=\frac{B^{3}}{9} .
\label{energy}
\eea

\section{Quantization of the system and eigen-functions}

We turn now to quantize our dynamical system, proceeding as
outlined in Sec. \ref{sec:Intro}. Since the near-singularity dynamics of
$z$ is trivial, we shall focus here on the quantization of $R$
(the trivial quantum evolution of $z$ will be briefly addressed at
the end of Sec. \ref{sec:time}).
The dynamics of $R$ is thus described by a wave-function
$\Psi(R,t)$, which admits the usual probabilistic interpretation
of Schrodinger theory. The wave
function evolves in time according to the Schrodinger equation
\begin{align}
i \hbar\partial_{t} \Psi=\hat H \Psi.\label{Schrodinger}%
\end{align}
Here $\hat H$ denotes the Hamiltonian operator, namely an operator
version of the semiclassical Hamiltonian
(\ref{Hamiltonian_Classic}), in which $P$ is replaced by the
momentum operator $\hat P=-i \hbar\partial_{R}$.

Since $\hat P$ is a differential operator, the ordering of
$\hat P$ and the factor $1/R$ matters, and different orderings may lead to
different quantum theories. There are many possible orderings, and a priori it
is not clear which is the "right" one. Here we choose the simplest symmetric ordering:
\begin{align}
\hat H = \hat P\frac{1}{R}\hat P = - \hbar^{2} \partial_{R} \frac{1}{R}
\partial_{R} \, .\label{H_quantum}%
\end{align}

In 1+1 dimensions, the combination $\hbar G/c^{3}$ turns out to be
dimensionless (for any choice of length, time, and mass units).
Since we have already chosen
General-Relativistic units $c=G=1$, the constant $\hbar$ is
dimensionless. In what follows we choose the value $\hbar=1$. A
different choice of $\hbar$ will amount to a rescaling of the
"Energy" parameter $E$ (which we shortly introduce)---and correspondingly a
rescaling of the time scale for quantum evolution.
\footnote{As was
already mentioned above, our effective Hamiltonian $H$ is
determined at the semiclassical level up to an arbitrary
multiplicative constant. Changing this constant will also result
in a rescaling of $E$ and of the time-scale for quantum evolution.}

The system's eigenstates $\psi_{E}(R)$ are determined by the time-independent
Schrodinger equation $\hat H \psi_{E}(R) =E \psi_{E}(R)$, namely
\begin{align}
\left( \frac{d}{dR}\, \frac{1}{R}\, \frac{d}{dR}\right) \psi_{E}(R)=
-E\psi_{E}(R).
\end{align}
 The general solution of this equation is
\begin{align}
\psi_{E}(R) = C_{A} Ai^{^{\prime}}(-E^{1/3}R) + C_{B} Bi^{^{\prime}}%
(-E^{1/3}R)\label{eigen_modesAB}%
\end{align}
(for $E\neq0$), where $Ai$ and $Bi$ respectively denote the Airy
functions of the first and second kinds, the prime denotes a
derivative with respect to their argument, and $C_{A}$,$C_{B}$ are
arbitrary coefficients. \footnote{For negative $E$, by $E^{(1/3)}$
we actually refer to the real branch $-|E|^{(1/3)}$. Note also
that for $E=0$ the solution is trivial, $\psi _{E=0}(R) = C_{1} +
C_{2} R^{2} $. However, since the spectrum is continuous, the
single exceptional eigenvalue $E=0$ will have no special
significance.}

Note that both functions $Ai(x)$ and $Bi(x)$ are real, and are analytic at
$x=0$. It is remarkable that unlike the semiclassical singularity at $R=0$ ---
and despite the singular $1/R$ factor which explicitly appears in the
Hamiltonian --- all eigenfunctions $\psi_{E}(R)$ are absolutely regular at
$R=0$. This fact is crucial to our approach: It implies that a generic wave
packet $\psi(R,t)$ propagating towards small $R$ will not experience any
singularity on approaching $R=0$, hence it will admit a unique and
well-defined evolution afterward.

We may only use eigenfunctions which are square-integrable. Note the far-zone
asymptotic behavior of the Airy functions: For $x>>1$ we have
\begin{align}
& Ai(x)\approx\frac{1}{2\sqrt{\pi}x^{1/4}}\,\,exp(-\frac{2}{3}x^{3/2}),\\
& Bi(x)\approx\frac{1}{\sqrt{\pi}x^{1/4}}\,\,exp(\frac{2}{3}x^{3/2}),
\end{align}
and on the negative axis:
\begin{align}
& Ai(-x)\approx\frac{1}{\sqrt{\pi}x^{1/4}}\,\,sin(\frac{2}{3}x^{3/2}+\frac{1}%
{4}\pi),\\
& Bi(-x)\approx\frac{1}{\sqrt{\pi}x^{\frac{1}{4}}}\,\,cos(\frac{2}{3}x^{3/2}%
+\frac{1}{4}\pi).
\end{align}
At the side of positive $x$ (negative $ER$),
the $Bi^{^{\prime}}$ eigenfunctions
diverge exponentially in $|R|^{3/2}$, and are hence non-normalizable.
On the other hand, the functions $Ai^{^{\prime}}(x)$ admit
a well-behaved asymptotic behavior for both signs of $R$
(exponential decay at one side and oscillations at the other side).
Thus, the permitted eigenfunctions are
\begin{equation}
\psi_{E}(R)=Ai^{^{\prime}}(-E^{1/3}R),\label{eigen_modes}%
\end{equation}
yielding an unbounded, non-degenerate, continuous spectrum.

In the next section we shall construct a well-localized initial
wave packet far from the singularity \footnote{By "far from the
singularity" we mean that the wave-packet distance (in $R$) from
the singularity is much larger than its width (namely, $r_0 >>
\delta r_0$ in the notation introduced below). Yet, as explained
in Secs. I and II, the region of spacetime addressed in this paper
is the asymptotic region near the singularity, namely $|R|<<K$.
These two assumptions do not conflict, provided that $\delta
r_0<<K$.}, and the far-zone asymptotic behavior of the
eigenfunctions will play a key role in the construction. We
introduce new variables related to $R$ and $E$, in order to give
this asymptotic behavior a more standard form of harmonic
functions.

First we define $r(R)$ (and its inverse) to be
\begin{equation}
r \equiv \frac{\sqrt{2}}{3}\,sign(R)\, |R|^{3/2}\, , \,\,\,
R \equiv \frac{3^{2/3}}{2^{1/3}}\,sign(r)\, |r|^{2/3}
.\label{kr_definitions}%
\end{equation}
We also define a variable $k$ satisfying
\begin{equation}
E=\frac{1}{2}k^{2} \, .\label{k_deifinition}%
\end{equation}

We shall only use $E>0$ eigenfunctions for the construction of our wave packet
(see next section).
In terms of the new variables $k,r,$ the $E>0$ eigenfunctions
$\psi_{k}(r)\equiv\psi_{E(k)}(R(r))$ read
\begin{equation}
\psi_{k}(r)=Ai^{^{\prime}}[-sign(r)\,|(3/2)k\,r|^{2/3}].\label{kr_eigenstates}%
\end{equation}
Their far-zone asymptotic behavior becomes:
\begin{equation}
\psi_{k}(r)\approx - \frac{|(3/2)k\,r|^{1/6}}{\sqrt{\pi}}\,\cos({|k|r}%
+\pi/4)\quad\quad(|k|r>>1)\label{kr_positive}%
\end{equation}
and
\begin{equation}
\psi_{k}(r)\approx-\frac{|(3/2)k\,r|^{1/6}}{2\sqrt{\pi}}\,exp(-|k\,r|)\quad
\quad(|k|r<<-1)\,.\label{kr_negative}%
\end{equation}

\section{Constructing the initial wave packet}

Next we construct a well-localized initial wave packet $\Psi_{0}(r)\equiv
\Psi(r,t=0)$ from the system's eigenfunctions. In the analogous free-particle
problem, it is often found convenient to consider gaussian wave packets.
Inspired by the harmonic form of the asymptotic behavior (\ref{kr_positive})
(along with the relation $E\propto k^{2}$), which resembles a free particle,
we shall seek an approximately-gaussian initial wave function in our problem
too. This gaussian is presumably centered at some initial location $r_{0}>0$,
with width $\delta r_{0}$, and it propagates from large to small $r$ with a
certain averaged wave-number $k_{0}<0$. The desired fiducial gaussian is thus
\begin{equation}
\sim \,e^{i\,k_{0}\,r}\,\,e^{-\frac{(r-r_{0})^{2}}{\delta r_{0}^{2}}}.\label{fiducial}%
\end{equation}

We design our initial wave function to be sharply peaked in both $r$-space and
$k$-space. Correspondingly, we demand $\delta r_{0}<<r_{0}$ and $\delta
k<<|k_{0}|$ , where $\delta k\equiv1/\delta r_{0}$. These inequalities (which
may always be achieved by choosing sufficiently large $r_{0}$ and $|k_{0}|$)
also imply $r_{0}|k_{0}|>>1$.

To accord with the above requirements, we take our initial wave function to be
\begin{equation}
\Psi_{0}(r)=\int_{-\infty}^{\infty}c(k)\,\psi_{k}%
(r)\,dk,\label{decomposition_initial}%
\end{equation}
where
\begin{equation}
c(k)=\beta e^{-i\,r_{0}\,\tilde{k}}\,\,e^{-\delta r_{0}^{2}\,\tilde{k}^{2}/4},\label{c_k}%
\end{equation}
$\tilde{k}\equiv k-k_{0}$, and $\beta$ is a certain normalization factor (determined below).
\footnote{In constructing this $\Psi_{0}(R)$ we have
only used $E>0$ modes. Note that
the modes of negative
$E$ oscillate at $r<0$ but decay exponentially at $r>0$. These
modes are inappropriate for representing our initial wave packet,
which is required to be sharply peaked at positive $r\sim r_{0}$.}
Note that $c(k)$ is just the $k$-th Fourier coefficient associated
with the fiducial gaussian (\ref{fiducial}). Figure
\ref{fig_Localized} below illustrates (for the specific case
$r_{0}=4, k_{0}=-40, \delta r_{0}=1/2$) that this $\Psi_{0}(R)$
has the desired well-localized gaussian-like shape.

We now analytically demonstrate that Eqs. (\ref{decomposition_initial},\ref{c_k}),
combined with the above inequalities, indeed yield an approximate Gaussian.
Since our wave packet is presumably sharply
peaked in both $r$-space and $k$-space, the inequality $r_{0}|k_{0}|>>1$
ensures that $|k|\,r>>1$ is satisfied throughout the phase-space region which
dominates $\Psi_{0}(r)$. We may therefore approximate the eigenfunctions
$\psi_{k}(r)$ by their far-zone harmonic form (\ref{kr_positive}). Also, in
the factor $\cos({|k|r}+\pi/4)$ therein we may replace ${|k|}$ by $-k$ to
facilitate analytic integration.
\footnote{This leads to a negligible error
because for all eigenfunctions with $k>0$, $c(k)$ is suppressed by the factor
$\exp[(k_{0}\ \delta r_{0}/2)^{2}]>>1$.}
Furthermore, in the prefactor $|(3/2)k\,r|^{1/6}$
we may approximate $k$ by $k_{0}$.
We obtain the approximate asymptotic form
\begin{equation}
\psi_{k}(r)\approx \alpha |r|^{1/6} \cos(-{kr}+\pi/4) ,
\label{kr_positive_aprx}
\end{equation}
where $\alpha \equiv -\pi^{{-1/2}}{(\frac{3}{2}|k_{0}|)^{1/6}}$.
Substituting this approximate $\psi_{k}(r)$ in
Eq. (\ref{decomposition_initial}) and integrating over $k$, one
obtains
\begin{equation}
\gamma_{0} \, |r|^{1/6}  \left[  e^{i\,k_{0}\,r}e^{-\frac{(r-r_{0})^{2}}{\delta r_{0}^{2}}%
}-i\,e^{-i\,k_{0}\,r}e^{-\frac{(r+r_{0})^{2}}{\delta r_{0}^{2}}}\right]
,\label{doublet0}%
\end{equation}
where $\gamma_{0}=-\frac{(-1)^{3/4}\sqrt{\pi}}{\delta r_0} \alpha \beta $ .
This describes the desired gaussian at $r\sim r_{0}$ plus
an undesired one at $r\sim-r_{0}$. One should recall, however,
that the harmonic approximation (\ref{kr_positive_aprx}) for
$\psi_{k}(r)$ only holds for positive $r$ (with $|k|r>>1$). For
negative $r$ the eigenfunctions are instead exponentially damped
(by the huge factor $exp(|k_{0}|r_{0})>>1$), as seen in Eq. (\ref{kr_negative}).
Therefore we are only left with the desired positive-$r$ (approximate
\footnote{Note that over the narrow gaussian (recall $\delta r_{0}<<r_{0}$), the factor
$|r|^{1/6}$ may be approximated as a constant, and we may therefore refer to Eq. (\ref{init_approx})
as an approximate gaussian.
The same applies to the moving gaussian of $\Psi(r,t)$ (in stages 1 and 3), considered in Sec.~\ref{sec:time} .
}) gaussian:
\begin{equation}
\Psi_{0}(r)\approx \gamma_{0} \, |r|^{1/6} \, e^{i\,k_{0}\,r}e^{-\frac{(r-r_{0})^{2}}{\delta
r_{0}^{2}}}.
\label{init_approx}
\end{equation}
The normalization condition $\int|\Psi_{0}|^{2}(dR/dr)dr=1$ now implies
$\gamma_{0}=\frac{3^{1/6}}{2^{1/12}\sqrt{\delta r_{0}
}\,\pi^{1/4}}$, which in turn yields
$\beta=-\frac{(1+i)\sqrt{\delta
r_0}}{2^{5/12}\pi^{1/4}|k_0|^{1/6}}$ .

Summarizing this section, our initial wave function $\Psi_{0}(r)$ is given by
Eq. (\ref{decomposition_initial}) combined with Eqs. (\ref{c_k})
and (\ref{kr_eigenstates}) --- which yield the desired well-localized, approximately-gaussian
shape (\ref{init_approx}).
Switching back from $r$ to $R$ we obtain
\begin{equation}
\Psi_{0}(R)=\beta\int_{-\infty}^{\infty}\,e^{i\,r_{0}\,\tilde{k}-\delta
r_{0}^{2}\,\tilde{k}^{2}/4}\,Ai^{^{\prime}}[-\frac{|k|^{2/3}}{2^{1/3}}%
R]\,dk\,,\label{final_gausian _R}%
\end{equation}
where, recall, $\tilde{k}=k-k_{0}$.

Figure \ref{fig_Localized} displays both the exact $\Psi_{0}(R)$
of Eq. (\ref{final_gausian _R}) and its approximate form
(\ref{init_approx}) (with $r$ translated back to $R$), for the set
of parameters $r_{0}=4$, $k_{0}=-40$, and $\delta r_{0}=1/2$. The
two graphs are indistinguishable, indicating the validity of the
above approximations.

\begin{figure}[h]
\begin{center}
\includegraphics[scale=0.85]{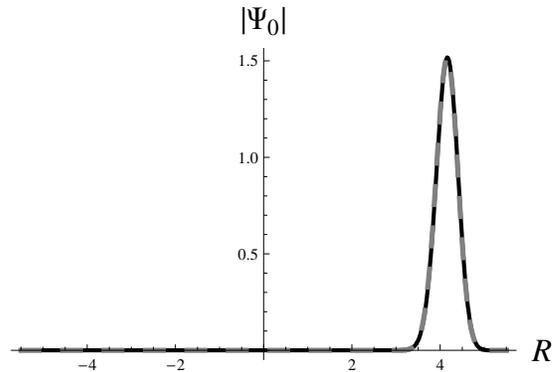}
\end{center}
\caption{The initial wave function $|\Psi_0|$ as a function of $R$,
constructed from Eq. (\ref{final_gausian _R}) (solid black), and compared with the approximation (\ref{init_approx}) (dashed gray).
Note the non-existence of the "mirror" Gaussian on the negative side of $R$.
Note also that $r_{0}=4$ corresponds to initial location $R= 3^{2/3}2\approx 4.16$ .}%
\label{fig_Localized}%
\end{figure}

\section{Time evolution of the wave packet}\label{sec:time}

The time evolution of $\Psi$ is dictated by the Schrodinger equation
(\ref{Schrodinger}), which endows each component $\psi_{k}$
in Eq. (\ref{decomposition_initial})
with an extra phase factor $\exp(-iEt)$:
\begin{equation}
\Psi(r,t)=\int_{-\infty}^{\infty}\,c(k)\,\psi_{k}(r)\,{{e^{-i(\frac{1}%
{2}k^{2})t}}}\,dk.\label{decomposition_evolution}%
\end{equation}
More explicitly, transforming $r$ back to $R$,%
\begin{equation}
\Psi(R,t)=\beta\int_{-\infty}^{\infty}\,{{e^{-i(\frac{1}{2}k^{2})t}}%
}\,e^{-i\,r_{0}\,\tilde{k}-\delta r_{0}^{2}\,\tilde{k}^{2}/4}\,Ai^{^{\prime}%
}[-\,\frac{|k\,|^{2/3}}{2^{1/3}}R]dk.
\label{wave_packet_construction_time}
\end{equation}
We shall first inspect $\Psi(R,t)$ numerically, and then explore it analytically.

\subsection*{Numerical evaluation}

We numerically evaluated Eq.~(\ref{wave_packet_construction_time})
for the particular case $r_{0}=4$, $k_{0}=-40$, $\delta r_{0}=1/2$
(yielding $\delta k\equiv1/\delta r_{0}=2$).

To this end we approximate the integral over $k$ by a summation
over a discrete set of equally-spaced values of $k$, separated by
a small interval $\varepsilon$, in the range $k_{0}-\Delta k\geq
k\geq k_{0}+\Delta k$. To achieve numerical precision the
parameters $\varepsilon,\Delta k$ must satisfy
$\varepsilon<<\delta k$ and $e^{-(\Delta k/2\delta k)^{2}}<<1$. We
choose here $\varepsilon=\delta k/20=0.1$ and $\Delta k=5\delta
k=10$. The plots were insensitive to any further increase of
$\Delta k$ or decrease of $\varepsilon$. \footnote{Changing from
integration over $k$ to sum over a discreet set of $k$ values
results in an infinite sequence of spurious replica gaussians,
with $r$-separation that scales as $1/\varepsilon$ (like in
standard Fourier decomposition). Our $\varepsilon$ is sufficiently
small that all spurious copies are well separated and do not
interfere with the authentic function $\Psi$ that we display.}\

\begin{figure}[htb]
\includegraphics[scale=0.7]{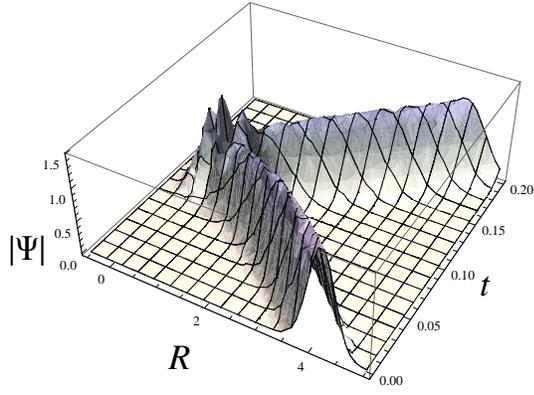}\
\caption{A three-dimensional plot describing the time evolution of
the wave function $|\Psi(R,t)|$.
The bounce at $R=0$ is evident.} \label{fig_3Dplot}
\end{figure}

\begin{figure}[h]
\subfigure[]{\includegraphics[scale=0.5]{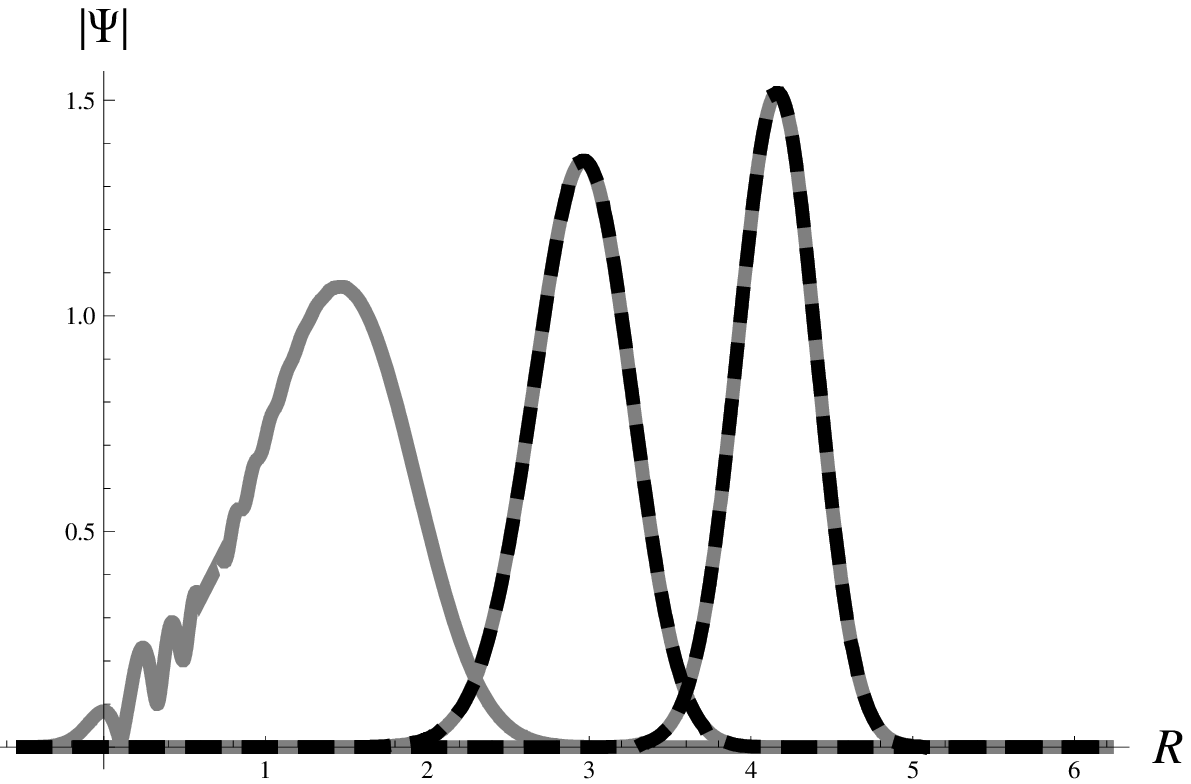}}\
\subfigure[]{\includegraphics[scale=0.5]{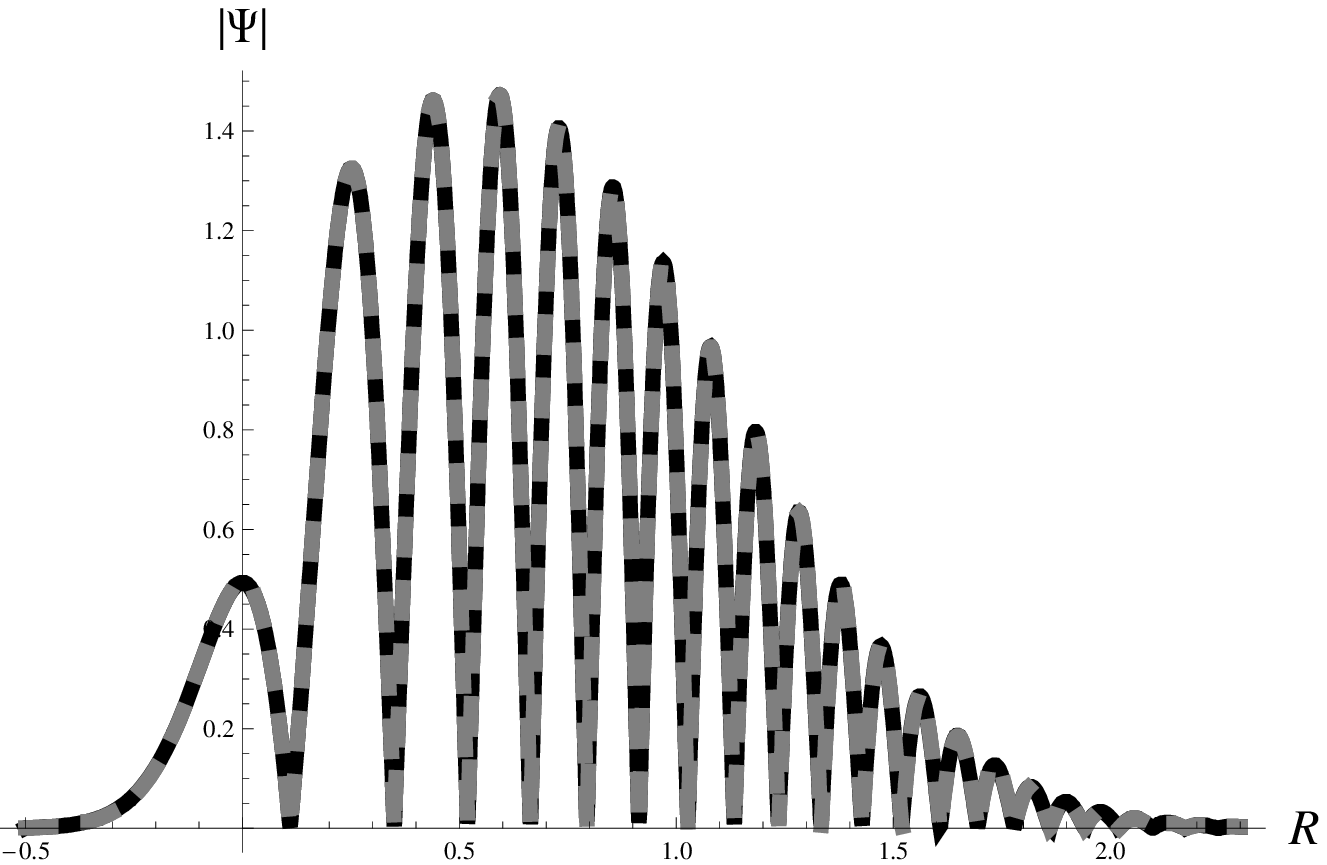}}
\subfigure[]{\includegraphics[scale=0.5]{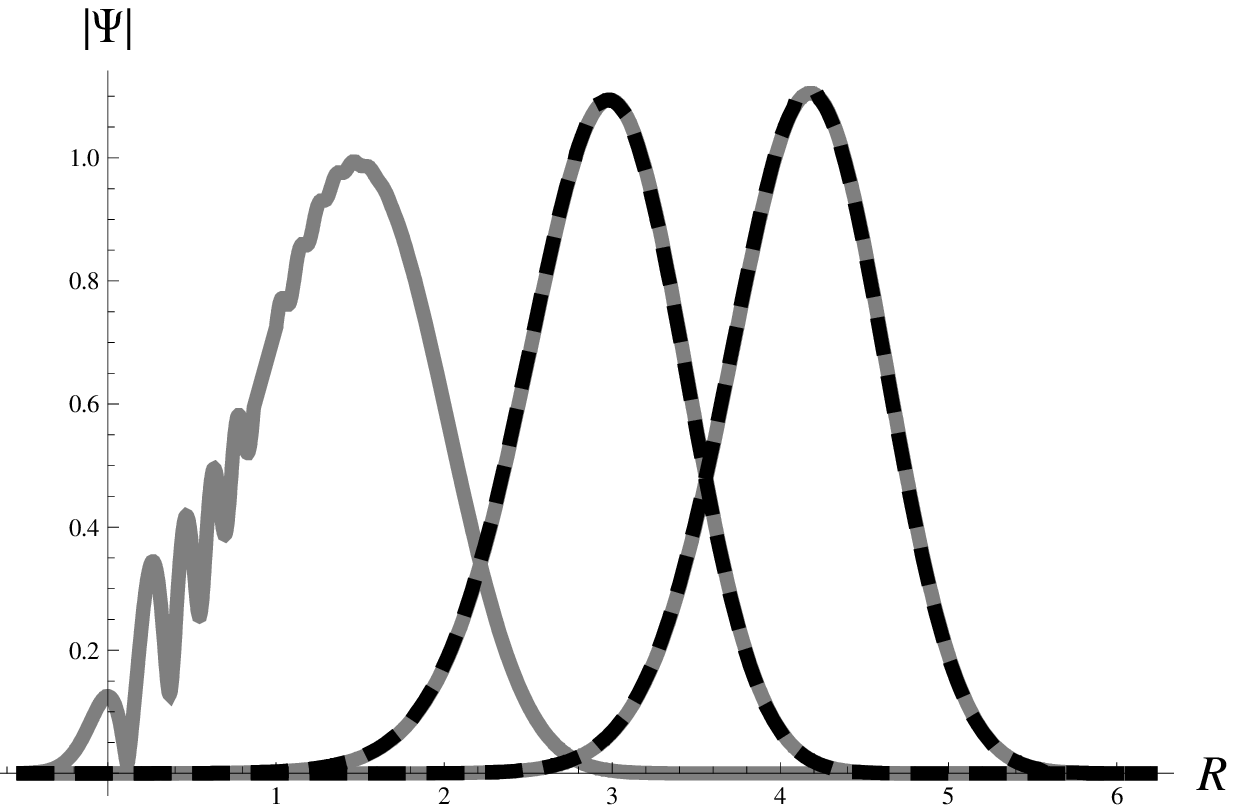}} \caption{ Plots
of $|\Psi|$ as a function of $R$ at several different times, as it
propagates towards the singularity (a), hits the singularity (b),
and then recedes from it (c). The solid-gray lines and
dashed-black lines represent the numerical calculations and
analytic approximations, respectively. Figure 3a displays three
snapshots which, going from right to left, correspond to times
$t=0,\,0.04,\,0.08$. Fig.~3b Displays $|\Psi|$ at $t=t_{hit}=0.1$.
Fig.~3c again displays three snapshots which, going from left to
right, correspond to times $t=0.12,\,0.16,\,0.2$. The analytic
approximation (dashed-black line) in the two right-most graphs of
both figures 3a and 3c corresponds to Eq. (\ref{right_gaussian}).
In Fig.~3b the analytic approximation corresponds to Eq.
(\ref{t_hit_approx}).} \label{fig_Dinosaur_Evolution}
\end{figure}

Figure \ref{fig_3Dplot} shows a three-dimensional plot of
$\Psi(R,t)$. It clearly demonstrates how the incidenting gaussian
bounces off $R=0$. The time evolution roughly consists of three
stages: (1) propagation from positive $R$ toward $R=0$, (2) the
intermediate stage, near the moment of bounce, and (3) propagation
of the wave packet back from the neighborhood of $R=0$ towards
large positive $R$.
On approaching the singularity (stage 2) the wave packet develops
wiggles, but subsequently it bounces and becomes nicely-peaked
again (stage 3). This behavior is seen more clearly in Fig.~
\ref{fig_Dinosaur_Evolution}, which displays a few snapshots of
$\Psi(R,t)$.

We also verified numerically that the wave function preserves its
unit norm upon evolution as expected. \footnote{Over the time
range covered by Fig.~\ref{fig_Dinosaur_Evolution}, the norm was
numerically found to vary by less than $10^{-3}$. }


\subsection{Analytic study}

Suppose that at some moment the wave function is located away from
$r=0$, in a range of $r$ satisfying $r\,|k_{0}|>>1$. Then, as was
discussed in the previous section, the oscillatory far-zone
asymptotic behavior applies and we may represent the
eigenfunctions by their approximate form (\ref{kr_positive_aprx}).
It is useful to define an auxiliary function $\tilde{\Psi}(r,t)$
by substituting this approximation for $\psi_{k}(r)$ in Eq.
(\ref{decomposition_evolution}):
\begin{equation}
\tilde{\Psi}(r,t)\equiv  \alpha \, |r|^{1/6}
\int_{-\infty}^{\infty}c(k)\,{{e^{-i(\frac{1}{2}k^{2})t}}}\,\cos
(-{kr}+\pi/4)dk.
\end{equation}
This function provides valuable insight into the dynamics of
$\Psi$ as we shortly demonstrate. The integration is
straightforward [in fact this integral is exactly the same as the
time evolution of a free Schrodinger particle with a
double-gaussian initial state as in Eq.~(\ref{doublet0})]. We
present $\tilde{\Psi}$ as \bea \tilde{\Psi}(r,t)=\gamma \,
|r|^{1/6} [
\Phi_{-}\,e^{i\,k_{0}\,r}\,e^{-\frac{[r-r_{c}(t)]^{2}}{\delta
r(t)^{2}}}\\ \nonumber
 +\Phi_{+}\,e^{-i\,k_{0}
\,r}\,e^{-\frac{[r+r_{c}(t)]^{2}}{\delta r(t)^{2}}}]  ,
\label{double} \eea where
$\gamma=\frac{3^{1/6}}{2^{1/12}\sqrt{\delta r(t)}\,\pi^{1/4}}$,
\begin{equation}
r_{c}(t)=r_{0}+k_{0}t ,
\label{peak_location}
\end{equation}
and
\begin{equation}
\delta r(t)=\sqrt  {\delta r_{0}^{2}+\left(\frac{2 t}{\delta
r_{0}}\right) ^{2}} \,  . \label{spread}
\end{equation}
$\Phi_{-}$ and $\Phi_{+}$ are certain phase factors
($|\Phi_{\pm}|=1$) whose explicit form will not concern us.
Evidently $\tilde{\Psi}(r,t)$ is a superposition of two gaussians
which propagate at opposite directions, one located at
$r=r_{c}(t)$ and the other at $r=-r_{c}(t)$. Both gaussians
approach $r=0$ at time
\begin{equation}
t_{hit}=-\frac{r_{0}}{k_{0}}.
\end{equation}

As long as the two gaussians comprising $\tilde{\Psi}$ are remote from $r=0$
--- namely, $|r_{c}(t)|>>\delta r(t)$ --- the inequality
$|r\,k_{0}|>>1$ is satisfied throughout both gaussians. Repeating
the argument which led to Eq. (\ref{init_approx}) above,
we find that the true wave-function
$\Psi$ will only contain the positive-$r$ Gaussian:
\begin{equation}
|\Psi(r,t)| \simeq \gamma  \, |r|^{1/6} \, e^{-\frac{[r-|r_{c}(t)|]^{2}}{\delta
r(t)^{2}}}\,\,\,\,\, \qquad |r_{c}(t)|>>\delta r(t) \, .
\label{right_gaussian}
\end{equation}

This single-gaussian shape characterizes the wave function in both
stages 1 and 3 (namely, the propagation toward $R=0$, and back
from $R=0$ towards large positive $R$, respectively). It is
illustrated in the two right-most graphs in both Figures
\ref{fig_Dinosaur_Evolution}a and \ref{fig_Dinosaur_Evolution}c.
These graphs display both the exact wave-function
(\ref{wave_packet_construction_time}) and the
(visually-indistinguishable) corresponding single-gaussian
approximation (\ref{right_gaussian}). Notice that the gaussians in
Fig.~\ref{fig_Dinosaur_Evolution}c are wider than their
counterpart in Fig.~\ref{fig_Dinosaur_Evolution}a, which simply
expresses the monotonic growth of $\delta r(t)$.

The intermediate stage 2 occurs at $t\sim t_{hit}$, when the
gaussian arrives at the very neighborhood of $r=0$. During this
stage $|\Psi|$ develops numerous wiggles, as demonstrated in
Fig.~\ref{fig_Dinosaur_Evolution}b (and also, to some extent, in
the left-most graphs of Figs.~\ref{fig_Dinosaur_Evolution}a and
\ref{fig_Dinosaur_Evolution}c). Qualitatively this behavior may be
interpreted as an interference pattern resulting from the overlap
of the two gaussians comprising $\tilde{\Psi}$. In particular the
right wings of both gaussians superpose at a range of positive $r$
values of order a few times $\delta r$. This superposition leads
to an oscillatory interference pattern in $\tilde{\Psi}$, with a
typical wave-number $k_{0}$. Since $|k_{0}|\delta r >>1$, the
far-zone harmonic approximation (\ref{kr_positive_aprx}) holds at
$r\sim\delta r>0$ (for both gaussians), hence
$\Psi\approx\tilde{\Psi}$ in this range. \footnote{On the other
hand, the $r<0$ portion of the interference pattern will not be
realized in $\Psi$, due to the exponential decay in the
corresponding far-zone asymptotic behavior (\ref{kr_negative}).}

This qualitative picture only yields a rough approximation for $\Psi$ in stage 2, because
near $r=0$ the far-zone approximation breaks. However, one can show (by
matching two different asymptotic approximations) that at the very moment of bounce
the following approximation holds:
\begin{equation}
\Psi(r,t=t_{hit})\cong \lambda e^{-q r^2}
Ai^{^{\prime}}[-(k_0^2/2)^{1/3}R(r)].
\label{t_hit_approx}
\end{equation}
where $q=\left(\delta r_0^2-2it_0\right)^{-1}$ and $\lambda$ is a certain coefficient.
Figure \ref{fig_Dinosaur_Evolution}b displays both this approximate expression
and the full function $\Psi(r,t=t_{hit})$, demonstrating a nice agreement.

Finally, we comment on the quantum dynamics of the other degree of freedom $Z$.
Since $Z$ evolves as a free particle [See Eq.~(\ref{RZ_dotdot})]
the quantum formulation adds nothing to its semiclassical dynamics:
The wave-function $\Psi_Z$, which describes the quantum
dynamics of $Z$, will just propagate freely, totally ignoring the
$R=0$ singularity.

\section{Semiclassical evolution beyond the singularity}

In both stages 1 and 3, the wave-function is sharply peaked at
\begin{equation}
r = |r_{c}(t)|=k_{0}|t-t_{hit}|.
\end{equation}
By the correspondence principle, this orbit should correspond to
some semiclassical solution. Indeed, translating back from $r$ to
$R$, we obtain
\begin{equation}
R(t)=B_{0}|{t-t_{0}|}^{\frac{2}{3}}, \label{correspondence}
\end{equation}
with $B_{0}=|3k_{0}/\sqrt 2|^{2/3}$ and $t_{0}=t_{hit}$, in full
consistency with Eq. (\ref{Solution_at_K_R}).

One of our main goals was to obtain the semiclassical extension
beyond the  $R=0$ singularity. This extension is now
determined (within the near-$R=0$ approximation) by the last
formula: It is just a time-reflection of the semiclassical
solution at $t<t_{0}$.

Note that the parameters $B_{0},t_{0}$ which characterize the
extension at $t>t_{0}$ are fully determined by the semiclassical
evolution at $t<t_{0}$. In particular, in Ref.
\cite{Dana_Amos_interior_design} it was found that for a
macroscopic ($M>>K$) CGHS black hole $B_{0} \approx [(3/2)\sqrt K
M]^{2/3}$, where $M$ denotes the (remaining) black-hole mass.

The other degree of freedom $Z$ just evolves steadily at $t \sim
t_0$ (at the leading order); Namely, \begin{equation} Z(t)\simeq
z_0 + z_1(t-t_0), \label{linear_Z}
\end{equation}
where $z_0,z_1$ are constants which are again determined by the
semiclassical evolution at $t<t_{0}$.

Note that the extension (\ref{correspondence}), which satisfies $B(t>t_{0})=B(t<t_{0})$,
is the only one which conserves $H$ across the singularity, as may be seen from Eq.~(\ref{energy})
(which holds at both $t<t_0$ and $t>t_0$).

It should be emphasized that the exact time-reflection symmetry
exhibited in Eq. (\ref{correspondence}) only characterizes the
\textit{leading-order} semiclassical evolution near $R=0$. Beyond
the leading order $R$ and $Z$ do couple [as may be seen, for
example, by substituting $S=(Z-\tilde R)/K$ in Eq.
(\ref{Rdotdot})], which spoils the time-reflection symmetry of
$R(t)$. The semiclassical evolution at $t>t_{0}$ {\it beyond} the
leading order may be obtained by taking Eq.
(\ref{correspondence}) and the above linear expression for $Z(t)$, translated back to
$\tilde R$ and $S$, as initial conditions for the full system
(\ref{Rdotdot},\ref{Sdotdot}).

\section{Discussion}

It is a common belief that Quantum Gravity will eventually cure the spacetime
singularities, which are known to occur in classical and semiclassical gravitational
collapse (as well as in cosmology). Here we demonstrated this idea by applying a
simplified quantization procedure to the spacelike $R=0$ singularity inside a
two-dimensional CGHS \cite{CGHS} evaporating BH. This allows us to explore the extension
of spacetime physics beyond the singularity.

Our quantization procedure is based on the minisuperspace approach, in which all
degrees of freedom associated with inhomogeneities are erased. The system's
evolution is then described by a wave-function, whose time evolution is dictated
by a Schrodinger-like equation, which we
solve analytically.

Our most crucial observation is that {\it quantum evolution is perfectly regular
(in fact, analytic) across the semiclassical $R=0$ singularity}. (This is far from
obvious, because the Hamiltonian operator itself is singular at $R=0$.) This
analyticity gives rise to unique quantum evolution across the singularity. We
find that an initially-localized wave packet spreads and fluctuates on approaching
$R=0$, but then it bounces back toward positive $R$ values. This is demonstrated
in Figs.~\ref{fig_3Dplot} and \ref{fig_Dinosaur_Evolution}.  Following the bounce,
the wave packet becomes sharply-peaked again, and approaches a well-defined
semiclassical orbit.

\begin{figure}[htb]
\includegraphics[scale=0.4]{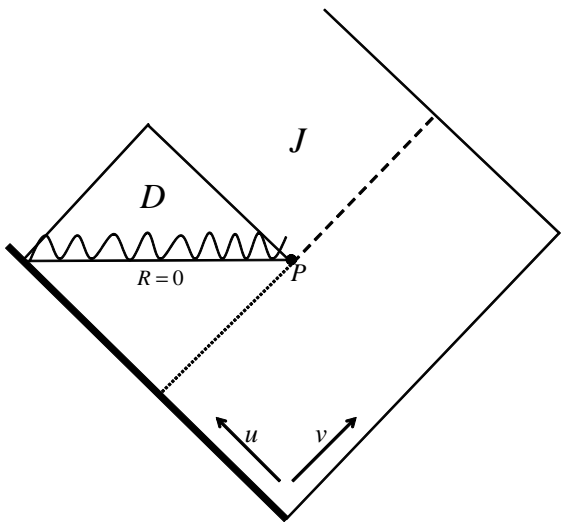}\
\caption{ Spacetime diagram of a CGHS black hole which forms by
gravitational collapse and subsequently evaporates. The thick null
line at the bottom left represents the collapsing massive shell.
The event horizon is marked by a dotted line. \textit{D}
represents the future Cauchy development of the singularity line
$R=0$. The domain marked by \textit{J} is discussed in the text.}
\label{fig_STD}
\end{figure}

The main objective of this paper was to explore the extension of
spacetime physics beyond the $R=0$ singularity inside the BH. The
wave-packet dynamics described above indicates that the strong
quantum fluctuations which develop at the very neighborhood of
$R=0$ quickly die out, giving rise to a new semiclassical phase
beyond the singularity. The quantum dynamics also determines the
structure of spacetime (namely, the spacetime dependence of  the
two semiclassical variables $R$ and $Z$) in this new patch, by
dictating the boundary conditions for $R$ and $Z$ across the
singularity: First, both variables are continuous. Second, $R$
bounces there at a (locally) time-symmetric manner, whereas $Z$
evolves steadily across the singularity. These boundary data for
both $R,Z$ and their time derivative (which in fact amount to
specifying the parameters $B_{0},t_{0},z_0,z_1$ of Sec. VI) give
rise to a unique and well-defined semiclassical Cauchy evolution
at the future side of the spacelike singularity, the region
denoted by $D$ in Fig.~\ref{fig_STD}

Interestingly, the situation found here resembles the recent developments in Loop
Quantum Cosmology (LQC) in many respects. Several analyses showed that in various homogeneous
cosmological models (open or closed universe, with or without cosmological constant) the
classical big-bang singularity is replaced in LQG by a regular bounce. \cite{list}
The same behavior (quantum singularity resolution, and a consequent
bounce) was obtained here in a two-dimensional BH---though within a more simplified quantum
formulation. There appears to be an important difference, though: In our case the very epoch
of bounce is marked by strong quantum fluctuations, whereas the cosmological bounce in LQC is
typically free of such strong fluctuations.

An obvious extension of this research is to obtain the specific
structure of semiclassical spacetime [namely, the functions
$R(u,v)$ and $S(u,v)$] beyond the singularity. In the present work
we obtained the required boundary data for these functions at the
future side of the singularity. The calculation of post-singular
$R(u,v)$ and $S(u,v)$ may naturally be divided into three stages:
(i) The asymptotic behavior in the very neighborhood of the
singularity is already given in Eqs. (\ref{correspondence}) and
(\ref{linear_Z}). It  applies as long as $R<<K$. \footnote{In
considering this stage of evolution [as well as stage (ii) below],
one should bear in mind that the approximate homogeneity of the
singularity only holds at spatial scales that are short compared
to the evaporation time scale. When considering the overall
large-scale post-singularity evolution, one should regard the
matching "parameters" $B_{0},t_{0},z_0,z_1$ as functions of $v$
which slowly drift along the line of singularity (just like the
effective local mass $M$). These four functions may be obtained by
evolving the CGHS field equations from the initial data up to the
(past side of the) singularity.} (ii) In an intermediate stage of
larger $t-t_0$ values, these simple asymptotic expressions no
longer hold. Furthermore, $R$ and $Z$ no longer evolve
independently. Still (as long as $t-t_0<<M/K$) the homogeneous
approximation applies, and the evolution of $R(t)$ and $S(t)$ is
determined by the ordinary differential equations
(\ref{Rdotdot},\ref{Sdotdot}). (iii) For even larger values of
$t-t_0$, the large-scale inhomogeneous character of spacetime
comes into expression, and the evolution of $R(u,v)$ and $S(u,v)$
must be determined by evolving the partial differential equations
(\ref{final_eq_simplified}). In both stages (ii) and (iii), the
evolution can be analyzed either numerically or by certain
analytic approximations (which are beyond the scope of the present
paper). A numerical simulation of the semiclassical CGHS field
equations is underway \cite{Liora,Pret}, and perhaps it will be
possible to extend this numerics to the post-singularity region as
well. By this way it should be possible to determine the
semiclassical spacetime [particularly the functions $R(u,v)$ and
$S(u,v)$] in the entire future "domain of dependence" of the
spacelike line $R=0$, the triangle-like region denoted by $D$
[which is actually $D_+(R=0)$] in Fig.~\ref{fig_STD}.

Note, however, that after the structure of the domain $D_+(R=0)$ is exposed, one will still face the
most challenging stage in the task of composing the overall post-singularity semiclassical spacetime:
Revealing spacetime physics in the "wedge" between  $D_+(R=0)$ and the external asymptotically-flat
region (the domain denoted $J$ in Fig.~\ref{fig_STD}). A priory it is not even clear if this "piece" is
connected (as suggested by the ATV point of view \cite{Ashtekar}, which assumes a fiducial Minkowskian
$\Re^{2}$ manifold) or disconnected (as may be suggested by the more common paradigm, see e.g.
Fig.~5 in Ref. \cite{Hawking}).  It seems that new concepts and insights will be needed in order to
explore this remaining piece of the puzzle.

\section{Acknowledgements}

We would like to thank Joseph Avron, Oded Kenneth, Martin Fraas,
and Joshua Feinberg for helpful discussions.  We are especially
grateful to Abhay Ashtekar for warm hospitality, for his advice
and encouragement and for many fruitful discussions. This research
was supported in part by the Israel Science Foundation (grant no.
1346/07).



\begin{thebibliography}{99}


\bibitem{Hawking_book}
See Sec. 8.2 in \textit{The Large Scale Structure of Space-Time}
S. W. Hawking and G. F. Ellis  (Cambridge university press 1973).


\bibitem{Hawking}
S. W. Hawking, Commun. Math. Phys. \textbf{43}, 199 (1975).


\bibitem{CGHS}
C. G. Callan, S. B. Giddings, J.A. Harvey and A. Strominger, Phys.
Rev. \textbf{D45}, R1005 (1992).


\bibitem{Singularity_first_notice}
J. G. Russo, L. Susskind and L. Thorlacius,
Phys. Lett. B\textbf{292}, 13 (1992).


\bibitem{Dana_Amos_interior_design}
D. Levanony and A. Ori, Phys.
Rev. \textbf{D80}, 084008 (2009).


\bibitem{tipler_singularity_strength}
J. Tipler, Phys. Lett. \textbf{64a}, 8 (1977)

\bibitem{Amos_Singularity_Strength}
See also A. Ori, Phys. Rev. D. \textbf{61} 064016 (2000).


\bibitem{Ashtekar}
A. Ashtekar, V. Taveras, and M. Varadarajan,
Phys.Rev.Lett.\textbf{100}, 211302 (2008).


\bibitem{list}
E. Bentivegna and T. Pawlowski, Phys. Rev. \textbf{D77}, 124025
(2008); A. Ashtekar, T. Pawlowski, P. Singh and K. Vandersloot,
Phys. Rev. \textbf{D75}, 024035 (2007); A. Ashtekar, T. Pawlowski
and P. Singh,  Phys. Rev. Lett. \textbf{96}, 141301 (2006); A.
Ashtekar, Gen. Rel. and Grav. \textbf{41} 707  (2009) and
references therein.

\bibitem{Liora}
L. Dori, research in progress

\bibitem{Pret}
F. M. Ramazanoglu and F. Pretorius, research in progress.


\end{thebibliography}
\end{document}